\documentclass[]{acmart}

\begin{document}

\title{\textit{ML-Quest}: A Game for Introducing Machine Learning Concepts to K-12 Students}

\author{Shruti Priya, Shubhankar Bhadra and Sridhar Chimalakonda}
\affiliation{%
  \institution{\\\textit{Research in Intelligent Software \& Human Analytics (RISHA) Lab} \\ Department of Computer Science and Engineering \\
  Indian Institute of Technology, Tirupati}
  \country{India}}
\email{Email: {cs18b043, cs18b034, ch}@iittp.ac.in}

% \author{Shubhankar Bhadra}
% \affiliation{%
%   \institutio     n{Indian Institute of Technology Tirupati}
%   \country{India}}
% \email{cs18b034@iittp.ac.in}
% \author{Sridhar Chimalakonda}
% \affiliation{%
%   \institution{Indian Institute of Technology Tirupati}
%   \country{India}}
% \email{ch@iittp.ac.in}

% \affiliation{%

% \author{Shruti Priya, Shubhankar Bhadra and Sridhar Chimalakonda 
% \affiliation{%
%  \textit{Research in Intelligent Software \& Human Analytics (\textit{RISHA Lab}})\\
%   Department of Computer Science and Engineering\\
%   Indian Institute of Technology Tirupati, India
%   }
% \email{cs18b043, cs18b034, ch@iittp.ac.in}}

\renewcommand{\shortauthors}{Shruti and Shubhankar, et al.}

\begin{abstract}
  Today, Machine Learning (ML) is of a great importance to society due to the availability of huge data and high computational resources. This ultimately led to the introduction of ML concepts at multiple levels of education including K-12 students to promote computational thinking. However, teaching these concepts to K-12 through traditional methodologies such as video lectures and books is challenging. Many studies in the literature have reported that using interactive environments such as games to teach computational thinking and programming improves retention capacity and motivation among students. Therefore, introducing ML concepts using a game might enhance students' understanding of the subject and motivate them to learn further. However, we are not aware of any existing game which explicitly focuses on introducing ML concepts to students using game play. Hence, in this paper, we propose \textit{ML-Quest}, a 3D video game to provide conceptual overview of three ML concepts: \textit{Supervised Learning}, \textit{Gradient Descent} and \textit{K-Nearest Neighbor (KNN) Classification}. The crux of the game is to introduce the definition and working of these concepts, which we call conceptual overview, in a simulated scenario without overwhelming students with the intricacies of ML. The game has been predominantly evaluated for its usefulness and player experience using the Technology Acceptance Model (TAM) model with the help of 23 higher-secondary school students. The survey result shows that around 70\% of the participants either agree or strongly agree that the \textit{ML-Quest} is quite interactive and useful in introducing them to ML concepts.
\end{abstract}


\begin{CCSXML}
<ccs2012>
   <concept>
       <concept_id>10003120.10003121.10003124.10010865</concept_id>
       <concept_desc>Human-centered computing~game for Learning</concept_desc>
       <concept_significance>300</concept_significance>
       </concept>
 </ccs2012>
\end{CCSXML}

\ccsdesc[300]{Human-centered computing~Game for Learning}


\keywords{Machine Learning, Games, Learning, K-12 Education}

\maketitle

\section{Introduction}
Machine Learning (ML) is pervasive today owing to the drastic increase in the availability of data and computational power \cite{von2020introducing}. ML has become relevant to a much broader community in recent years with an increase in addressing challenges across multiple domains such as job automation, privacy concerns, facilitate healthcare services and many more \citep{sulmont2019can, luccioni2021using}. It has been integrated into various products and services, from speech recognition system to movie recommendations \citep{von2020introducing, sharma2020systematic, nassif2019speech, lund2018movie}. However, the underlying process of ML is black-boxed for many users \citep{hitron2019can}.  %Considering the wide presence of ML and its impact on society, it has become important to raise awareness of ML knowledge to the people \cite{best10examples}. %In addition, given the inclusion of programming and computational thinking in curriculum for school students, it is also important to introduce foundations of ML concepts to school students for bringing awareness of this rapidly emerging technology.

Considering the vast presence of ML and its impact on society, it has become important for students, to be prepared for the upcoming digital era and empower themselves as digital citizens \citep{inproceedings}. %Today, students might have grown with technology in their hands, but having an introductory knowledge of this could help them to better understand the world \citep{passey2017computer}. 
Since past few years, computer science has become one of the significant part of primary and secondary education, as it promotes computational thinking \citep{k-12C}. Studies have reported that introducing computer science at an early stage could help students grow their interest in the subject, which could further aid the development of advanced technologies \citep{passey2017computer}. %  Introducing computational thinking helps students to formulate problems, find solutions and design systems based on computer science fundamentals \cite{inproceedings}.
% why we should teach ML in abstract and using non-conventional methods.
Along with conventional teaching methods such as books and video lectures, technologies such as Artificial Intelligence, Virtual Reality and Mixed Reality \citep{ARtutor,cooper2000alice,marques2012software} have been incorporated to introduce computer science concepts at the school level in order to promote computational thinking \citep{nashine2021generalized}. 

With the increasing popularity of ML due to its widespread usage and societal impact, there have been attempts to introduce ML as a K-12 subject through online courses and tutorials \citep{wallace2008integrating, mariescu2019machine,aki2021introducing}. ML is a field of computer science, which provides computers the ability to learn and solve unknown problems on their own \citep{sharma2018machine}. Uncovering the basic concepts of ML at an early stage could help students improve their cognitive thinking and could also prepare them for pursuing higher education and future career in ML \citep{von2020machine,hitron2019can}. Nevertheless, it should be noted that students should not be overwhelmed because of the complexity of an unknown concept \citep{hitron2019can, mariescu2019machine}. Therefore, there is a need to introduce ML concepts at an early stage without burdening them with inner complex details.

%Traditionally, teaching ML has been considered only for higher education such as undergraduate and PG courses \citep{torrey2012teaching}.
% Motivating the need to teach ML concepts using game
Overtime, many studies have shown the effectiveness of games in the field of learning which ultimately led to the involvement of them to teach multiple concepts across different domains such as code debugging \citep{venigalla2020g4d, lee2013game}, computer architecture \citep{tlili2016improving} and social awareness \citep{survivecovid}. Games have been considered to be entertaining and engaging \citep{sung2018facilitating}. Teaching concepts through games could help learners perceive learning as fun without overwhelming them with inner details of the concept \citep{eagle2009experimental,krajcsi2019algotaurus}. We believe that ML concepts can also be introduced via a game without going through the lower level of implementation details. However, based on the reported studies in the literature, we could not find any video game which explicitly focuses on teaching ML concepts.

Considering the effectiveness of games \citep{sung2018facilitating} and their wide presence in learning \citep{venigalla2020g4d, lee2013game, tlili2016improving}, we propose a 3D video game \textit{ML-Quest} which aims at introducing the definition and the working, which we call conceptual overview of three ML concepts: \textit{Supervised Learning}, \textit{Gradient Descent} and \textit{KNN Classification} to higher-secondary school students. The game aims to introduce the ML concepts to school students with no prior knowledge of ML, without overwhelming them with inner details such as underlying complex mathematics or any technical jargon. \textit{ML-Quest} comprises three levels to teach three different ML concepts. At each level player is equipped with clues, an instruction board, and dialogue messages to solve the task intuitively aligned to a particular ML concept.  At the end of each level, the underlying ML concept is defined and mapped with the task performed to help the player to understand the working of the concept. The proposed game aims to help K-12 students boost their understanding of ML at an early age by applying them in a simulated game world. %and also use these learned skills to solve problems faced on a daily basis.Contributions of this paper are as follows:The game aims to introduce the high level abstraction behind some of the selected ML concepts to higher secondary students and help cultivate interest among them for the subject, which could encourage them to opt ML as a future career option.

Evaluation of the game has been done considering its Ease of Use, Usefulness, Intention to use, and Correctness. Table \ref{tab:questions} summarizes the result of the survey following mean and standard deviation format. Result shows around 70\% of the participants either agree or strongly agree that the game has made learning more interactive with a mean of 3.86 and a standard deviation of 0.69. Participants also find the game, easy to use and very useful to get an overview of ML concepts with the average mean of all survey question related to ease of use and usefulness (refer Table \ref{tab:questions}) 3.61 and 3.78 respectively. Also, 70\% of the participants found the ML concepts taught at each level aligned to the gameplay, thus validates the correctness of the game. Results of the survey along with the game demo can be found here \footnote{\url{https://osf.io/czru5/?view_only=64741df07d0e481685e2dbb2afe7a1df}}

\section{Related Work}
% - game
% - games for computing
% - ML education for kids

% Over the decades, there have been several tools such as books, videos and demonstrations used for teaching computational and ML related concepts but many students find it difficult to understand these concepts and lack motivation to continue the subject further through these traditional methods of teaching \citep{nurjanah2018analysis}.

With advancements in technologies, several techniques such as visualisation, computational notebooks and game development have been adopted to make learning more interactive \citep{sulmont2019can,marques2012software}. Many studies have reported games to be helpful in introducing various educational concepts and it has been leveraged by many researchers in the field of learning \citep{lee2013game}. Games are motivating and make learning more interactive, interesting and engaging \citep{sung2018facilitating}. Teaching concepts through game could promote computational thinking and help students to understand the concept in a fun and better way.

\textit{Wu's Castle} \citep{eagle2008wu} is a 2D game to teach array and looping constructs. In this game, the player needs to write code to perform certain tasks and understand the concept though visualisation. The game has been evaluated on 28 students and results show that the students who have played the game outperformed in the coding test compared to those who did not play the game. Similarly, there are several games to teach concepts across different domains such as \textit{G4D} \citep{venigalla2020g4d} and \textit{Gidget} \citep{lee2013game} to help students in debugging codes, game to teach Computer Architecture to undergraduates \citep{tlili2016improving}, video game to teach supply chain and logistics \citep{liu2017using} and many more. All these games have been evaluated for their likelihood, usefulness and player experience, and results show that learning has become much easier and understandable though the games compared to the traditional way of learning for students.

Considering the societal impact of ML, there have been initiatives to include AI and ML into K-12 curriculum to promote computing education. Wangenheim et al. \citep{von2020machine} designed an online course for K-12 students to introduce basic ML concepts that also teaches to train ML model using Google Teachable Machine. Mariescu-Istodor et al. \citep{mariescu2019machine} proposes a method for teaching ML model training for object recognition to high school students based on the knowledge they had gained from their high school Mathematics and Computer Science classes. %But existing literature mostly focuses on programming and computational thinking \citep{lye2014review,da2018teaching}. 
However, the amount of literature or courses providing an overview on teaching ML to school students is minimal, mostly targeting undergraduate and postgraduate students which is evident from their surveys \citep{marques2020teaching}. 

There have been a few attempts to introduce ML concepts to K-12 through visualisation and tools. Chung et al. \citep{chung6introducing} proposes an approach to introduce basic concepts of ML to K-12 by making them develop applications to perform different tasks, using web based tools such as ``Machine Learning for Kids", ``Scratch 3" and ``Lego Mindstorms EV3 robots". Similarly, Christiane et al. \cite{von2020machine} introduces a course to promote machine learning among K-12 by guiding them to develop image recognition model using Google Teachable Machine. Hitron et al. \citep{hitron2019can} also tried to introduce ML concepts to the school children using hand gestures through a hardware device for training a classification model. From the outcomes, they conclude that if the black boxes are uncovered, it would be easier for children to understand the world around them better. They also mention that uncovering multiple layers of an unknown concept could overwhelm the students. So, there is a need for a balance between uncovering and black-boxing ML concepts. Although there have been attempts to introduce ML concepts to students, we could not find any interactive and engaging approach such as video games, which could introduce concepts to students without overwhelming them with the underlying complexities. Therefore, considering the advantage of games in learning, we propose \textit{ML-Quest}, a 3D video game to introduce ML concepts to school students without burdening them with any inner details (i.e underlying mathematics or any technical jargon).

\section{Design Methodology}

\textit{ML-Quest} is a 3D, Role Player Game (RPG) with a quest theme. The game has been designed keeping in mind the integration of educational topics to game play through challenging tasks aligned with the concept being taught, clear instructions and giving immediate feedback to the player. The game has a storyline where the protagonist is on a mission to protect her kingdom from the enemy referred as `red men'. The current prototype version of the game has three levels capturing three different ML concepts namely \textit{Supervised Learning}, \textit{Gradient Descent} and \textit{KNN Classification} through well-designed tasks. At each level the tasks have been designed by aligning to a particular ML concept and can be solved by carefully following the provided instructions (for details, refer section \ref{section:outcomes}). Each level is equipped with instruction boards and dialogue boxes to guide the player. In order to create a mental model of the underlying concepts, at the end of each level, the definition and working of the ML concept is displayed mapped with the tasks performed in the level.

Various games in the literature such as \textit{G4D} \citep{venigalla2020g4d} and \textit{Game for Computer Architecture} \citep{tlili2016improving} have used different game design patterns such as Scaffolding \citep{gonulal2018scaffolding} and Early Bird \citep{kelle2011design} to systematically address the recurring design problem and for effective game design. Scaffolding is observed to be positively bracing game development, which is evident with its wide usage in various educational games \citep{jantan2012experimental, venigalla2020g4d}. Hence, we have also opted for the scaffolding technique for \textit{ML-Quest}. Scaffolding in this context refers to hiding the high-level information and gradually uncovering the domain-specific concepts \citep{gonulal2018scaffolding}. We have implemented this idea by using metaphors for the tasks and processes at each level and unmasked these abstractions at the end of the level by defining and mapping them to the underlying ML concept.

According to Jantan et al., scaffolding consists of $10$ characteristics that lead to an effective game design \citep{jantan2012experimental}. In our game, we have tried to capture five of these characteristics: \textit{Provides clear direction}, \textit{Clarifies purpose}, \textit{Keep students on task}, \textit{Appropriateness of the instruction level} and \textit{Continuity}. The player is equipped with instructions and dialogues describing tasks and steps to complete each level. Thus, incorporating the scaffolding characteristics of \textit{Provides clear direction} and \textit{Clarify purpose}. The player performs tasks based on the instructions provided, which on completion is mapped to the underlying ML concept for a better understanding of the player, thus implementing the scaffolding characteristics of \textit{Keep students on task} and \textit{Appropriateness of the instruction level}. The game's complexity increases gradually based on the previously performed task, and the storyline progresses with a smooth transition between the levels, implementing the scaffolding characteristic of \textit{Continuity}.

\begin{figure}[!ht]
    \centering
    \includegraphics[width = 0.9\linewidth]{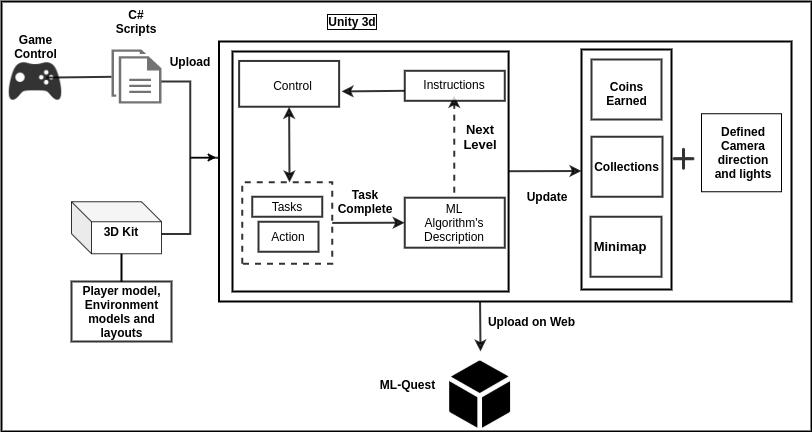}
    \caption{Architecture of \textit{ML-Quest}.}
    \label{fig:working}
    \vspace{-4mm}
\end{figure}

\subsection{Game Levels and their Outcomes}
% This section talks about the decisions taken for designing each level of the \textit{ML-Quest} incorporating ML concepts at each level.
% This section answers the question, \textit{how abstraction of the ML concepts has been implemented at each level of ML-Quest?}

The current version of the game has three levels, each introducing one ML concept namely \textit{Supervised Learning}, \textit{Gradient Descent} and \textit{K-Nearest Neighbor (KNN) Classification} respectively. In the broad scope of ML we have focused on the concepts those are fundamental and amenable \citep{ayodele2010machine} and hence in the prototype version primarily focused on these three concepts. The concepts and learning outcomes for each level have been discussed further.
%The levels have been designed to enforce the player to apply underlying high level idea of these ML concept intuitively by following some set of instructions, which are later revealed by mapping those applied concepts to the game element for better understanding of the player.%

\begin{figure*}[!ht]
  \includegraphics[width=\textwidth]{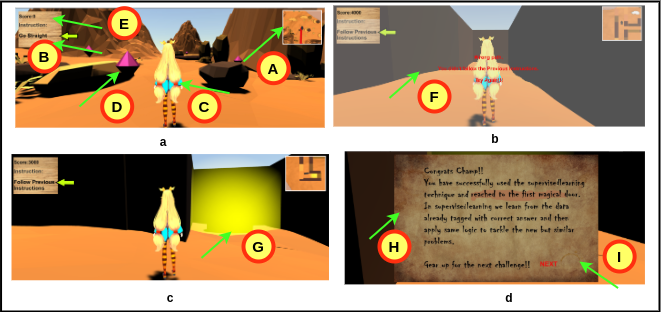}
  \caption{Scenes from \textit{Level 1}. Solving the Maze based on previously provided directions. [A] is minimap to see and follow red path, [B] shows instruction to follow, [C] is the player, [D] is diamond which needs to be collected to earn points, [E] tells the score, [F] warning message saying wrong path, [G] the destination (i.e magical door to next level), [H] displays learning outcome, [I] is the button t go to the next level.}
  \label{fig:level1}
\end{figure*}

\textit{Supervised Learning}: Level 1 of the \textit{ML-Quest} is designed to introduce the definition and working of \textit{Supervised Learning}. \textit{Supervised Learning} is a well-known ML technique in which the ML model is trained to find the solution based on previous problem-solution pair \citep{mohamed2017comparative}. This level simulates a similar scenario through a maze problem. Here, the player is considered as a model. With the help of provided instructions player needs to reach to the entrance of the maze (i.e. training the model) and then solve the maze by following the same previous directions (i.e. apply the previous knowledge to solve the new problem), as shown in Figure \ref{fig:level1}. An alert message pops up in case the player does not remember the previous instructions and tries to solve the maze by hit and trial, in this way, we make sure that the player is following previously taught instructions only. Once a player achieves the goal, the learning outcome is displayed, which states the definition of \textit{Supervised Learning} and then maps steps involved in the working of it to the task performed in the level so that player can get the conceptual overview of \textit{Supervised Learning}.

\begin{figure*}[!ht]
  \includegraphics[width=\textwidth]{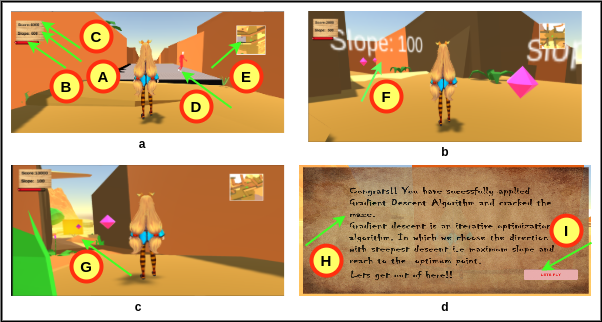}
  \caption{Scenes from \textit{Level 2}. Player reaches the magical door at the bottom of the maze using slope as a hint. [A] displays the slope of the following path, [B] is the health bar, [C] displays the score, [D] is the red man, [E] is the minimap displaying the possible directions, [F] displays slope of the path at intersection, [G] is the destination (i.e magical door to next level), [H] displays the learning outcome of the level, [I] is the button to the next level.}
  \label{fig:level2}
\end{figure*}

\textit{Gradient Descent}: \textit{Gradient Descent} is an optimization technique used to minimize a function by iterating in the direction of maximum slope and reach the local minima \citep{ruder2016overview}.
In Level 2, the player is instructed to apply the steps involved in the working of \textit{Gradient Descent} to solve the maze and reach the bottom-most point. The slope is used as a hint to choose the correct path (i.e., a path with maximum slope) and reach the optimum point as shown in Figure \ref{fig:level2}. After a player reaches the destination by choosing a path with maximum slope and overcoming all hurdles, the definition and working of the \textit{Gradient Descent} is introduced and mapped to the level for the player to easily grasp the concept. 

\begin{figure*}[!ht]
  \includegraphics[width=\textwidth]{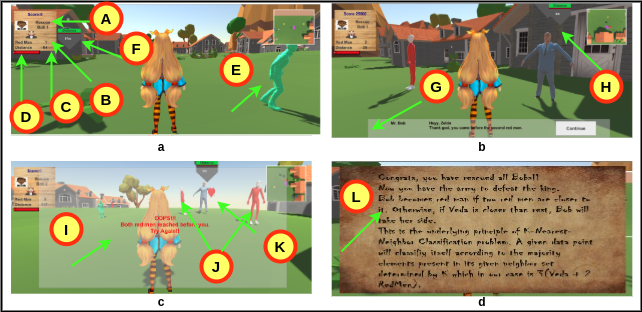}
  \caption{Scenes from \textit{Level 3}. Section [a], [b] shows player rescuing Bob before the two red men could reach and as a result learning outcome displays as shown in [d]. Section [c], shows the situation when both red men reaches before the player and hence player looses the level and it restarts. [A], [B], [C], [D], [E], [F] represents active Bob name which needs to be rescued, number of red men reached Bob, distance between Bob and red men, health bar, population, distance between player and Bob respectively. [I] displays the message when red men reached before the player, [L] is the learning outcome.}
  \label{fig:level3}
\end{figure*}

\textit{KNN Classification}: \textit{KNN Classification} is a ML technique used to classify objects based on the majority of elements among its k nearest neighbors \citep{mohamed2017comparative}. As the game is meant to provide an abstraction of the concept to the students, we have considered presenting this concept of majority in the form of persuasion power and voting. In Level 3, a player has two votes, and enemies have one vote each. Considering k as three, if the player reaches to target (a green-colored person) before the two enemies, then in the three closest neighbors of the target, the player will have a majority vote (i.e., two), so the target will be classified on the player's side. Alternatively, if two enemies reach before, they will be in the majority and will make the target to their side. A distance meter is provided, which shows the distance between player and target along with an enemy meter which, shows the distance between the enemy and the target. At the end of this level, the definition and working of \textit{KNN classification} mapped to the task performed in this level is displayed as shown in Figure \ref{fig:level3}. 
\label{section:outcomes}

\section{Development}
The game has been implemented using the Unity 3D game engine \footnote{\url{https://unity.com/solutions/console-and-pc-games}}. Figure \ref{fig:working} depicts the step-by-step approach used for developing the game. We have used Unity assets to implement different game objects in each level along with the environment. All the motions, listeners, and control of different game objects have been scripted in C\#. We have created minimap at each level for the user to get the top view of the surrounding using Unity, to render texture to the target texture field. Minimap also keeps track of the player position by dealing with the player position vectors and time variables. Player movement is controlled through input from the keyboard. A player is equipped with rich UI features such as a free-roaming environment, NPCs (Non-Playable Character), and dialogue boxes, which have been implemented considering the physics-oriented events and have been controlled with the help of object listeners and events of the game objects.

The game also provides navigation tools such as a distance-meter, path indicators, and instruction board to understand the gaming world better and reach the goal based on these perceptions. At the end of each level, the learning outcomes are displayed, attached with an object listener, to transition to the next level. The prototype of the game has been built in WebGL \footnote{\url{https://developer.mozilla.org/en-US/docs/Web/API/WebGL_API}} format and deployed on a website named \textit{Simmer.io} \footnote{\url{https://simmer.io}}, which is an online platform for sharing games.

\section{User Scenario}
Consider \textit{Drishti}, a school kid,  enthusiastic to learn about ML related concepts, decides to play the \textit{ML-Quest}. She chooses \textit{Devi} as her playable character and starts the game. After reading the instructions and storyline, she clicks \textit{Next} and level 1 begins. Here, \textit{Devi} is in an isolated desert as shown in Figure \ref{fig:level1}[a]. \textit{Devi} needs to follow the red path marked in the mini-map in order to reach the destination along the way, she needs to note down the direction of her movements displaying on the instruction board. After following the instruction set, she reaches the entrance of the maze as shown in Figure \ref{fig:level1}[a][A]. Now, \textit{Devi} follows the previously noted direction of her movement to solve the maze and reach the magical door as shown in Figure \ref{fig:level1}[c][G]. While crossing the maze, if \textit{Devi} steps into a wrong path, she is prompted with a warning as shown in Figure \ref{fig:level1}[b][F] and the level restarts. As soon as she reaches target by following all instructions and performing tasks carefully, a prompt appears with the learning outcome of level 1 and displays definition of \textit{Supervised Learning} and how she has used the steps involved in the working of \textit{Supervised Learning} without even knowing it, in order to achieve the goal as shown in Figure \ref{fig:level1}[d][H]. Clicking on \textit{Next} as shown in Figure \ref{fig:level1}[d][I] takes \textit{Devi} to level 2.

\textit{Devi} encounters a similar maze as shown in Figure \ref{fig:level2}.a. She needs to solve the maze by choosing the path with maximum slope value, as shown in Figure \ref{fig:level2}[b][F] To make the game more challenging, a few enemies (i.e., red men) randomly walk within the maze. Interaction with red men will reduce the health of the protagonist. If \textit{Devi} passes close to an enemy, her health decreases. In the worst case, if health becomes zero, the game ends, and the level restarts. Once \textit{Devi} reaches the second magical door as shown in Figure \ref{fig:level1}[c] by carefully choosing the path with maximum slope, the learning outcome is displayed as shown in Figure \ref{fig:level2}[d][H] which maps the level to steps involved in \textit{Gradient Descent} concept. Clicking on \textit{Next} as shown in Figure \ref{fig:level2}[d][I] will take her to the final level. %out of the desert into the final level.

\textit{Devi} finally reaches the town of Bobs as shown in Figure \ref{fig:level3}[a]. She needs to rescue three Bobs, one at a time, by getting his heart before the red men. If two red men, as shown in Figure \ref{fig:level3}[c][J] reach before the player, Bob will take the side of red men as shown in Figure \ref{fig:level3}[c] and the level restarts. Bob can be identified by a tag hovering over the character as shown in Figure \ref{fig:level3}[b][H]. \textit{Devi} starts searching the Bob with the help of distance-meter, shown in Figure \ref{fig:level3}[a][F], which displays the distance between the active Bob and \textit{Devi}. If \textit{Devi} finds Bob before the two red men, a dialogue appears, which tells about the next task as shown in Figure \ref{fig:level3}[b][G]. After the conversation ends, \textit{Devi} collects the heart similar to the one shown in Figure \ref{fig:level3}[c][K] and Bob comes to Devi's side, and she continues to search for the next Bob. Once all the three Bobs are rescued, a prompt, as shown in Figure \ref{fig:level3}[d][L] appears, indicating the learning outcomes which defines \textit{KNN Classification} and how the steps involved in the working of this concept has been implemented through the tasks performed by \textit{Devi} in this level.

\section{Evaluation}
\begin{table}
\caption{ Questions asked in survey along with their code}
  \label{tab:questions}
\resizebox{\columnwidth}{!}{\begin{tabular}{llll}
\toprule
\multicolumn{1}{c}{\textbf{Evaluation Questions}}                                                                   & \textbf{Variable} & \textbf{Mean} & \textbf{SD} \\ \midrule
I think learning to use ML-Game is easy                                                                             & EU1           & 3.52          & 0.95        \\
I think becoming skillful at using ML-Game is easy                                                                  & EU2           & 3.70          & 0.93        \\
Using ML-Game would improve my understanding of the concepts in Machine Learning.                                   & U1            & 3.91          & 0.85        \\
Using ML-Game would enhance my effectiveness in learning and understanding multiple principles of Machine Learning. & U2            & 3.78          & 0.74        \\
Using ML-Game would make it easier for me to learn Machine Learning concepts.                                       & U3            & 3.65          & 0.78        \\
Assuming I had access to ML-Game, I intend to use it.                                                               & I1            & 3.43          & 0.95        \\
ML-Game has made my learning interactive                                                                            & I2            & 3.87          & 0.69        \\
Visualizations displayed by ML-Game are relevant to the concept taught at each level.                               & C1            & 3.65          & 0.83        \\
\bottomrule
\end{tabular}}
\end{table}

\begin{table}
\caption{Demographic questions asked in survey}
{\begin{tabular*}{\textwidth}{@{\extracolsep{\fill}}ll} \toprule
\multicolumn{1}{c}{\textbf{Demographic Questions}} & \textbf{Percent (\%)} \\ \midrule
\textbf{Prior knowledge about ML}                  &                       \\
Yes                                                & 33.3                  \\
No                                                 & 66.7                  \\
\textbf{Inquisitiveness for ML}                    &                       \\
Yes                                                & 42.1                  \\
No                                                 & 57.9                  \\
\textbf{Choice of the platform}                    &                       \\
Laptop/Desktop                                     & 69.6                  \\
Smartphone                                         & 47.8                  \\ \bottomrule
\end{tabular*}}
\label{tab:demographics}
\end{table}

\textit{ML-Quest} is a 3D quest theme based game, meant to introduce ML concepts to students through interesting storyline and engaging tasks. Considering these points, \textit{ML-Quest} has been primarily evaluated for its \textit{Ease of use (EU)}, \textit{Usefulness (U)} and \textit{Intention to Use (I)}.

As per the literature, many games \citep{venigalla2020g4d, lee2013game, hakulinen2011card, tlili2016improving} have been evaluated using a questionnaire-based user survey combined with TAM model. TAM has been widely used over time to evaluate learning games such as multimedia learning environment \citep{saade2007viability}, ARTutor \citep{ARtutor}, and many more. The two main beliefs of TAM model, which influences the decisions taken by a user for adopting a new technology are \textit{Perceived Ease of use (EU)} and \textit{Perceived Usefulness (U)} \cite{davis1989perceived}. Considering the current scope of \textit{ML-Quest} and factors associated to the TAM model, we found that TAM fits well for the evaluation of our game. Along with these two factors TAM model also includes factors such as \textit{Privacy concern}, \textit{Perceived risk}, \textit{Facilitating conditions} and \textit{Subjective norm} \cite{TAM1}. But since we are not dealing with any confidential user data, checking for \textit{Privacy concern} and \textit{Perceived risk} would be reluctant in case of our game. Since, \textit{ML-Quest} is a video game specially for K-12 students and can be played on any browser without need of any supervision, evaluating for \textit{Facilitating conditions} and \textit{Subjective norm} is also not useful. So, we have excluded these factors and considered only the above mentioned two relevant factors (i.e \textit{Perceived Ease of use (EU)}, \textit{Perceived Usefulness (U)}). \textit{Intention to Use (I)} and \textit{Correctness (C)} are also the important factors as \textit{Intention to Use}, tells whether the user will be willing to use the game in future or not and \textit{Correctness (C)}, defines the degree of validity of the game, which is relevant in our case as we want to evaluate the relevance of the tasks assigned in each level compared to the concept taught. Thus, we are evaluating our game based on four factors (i.e \textit{Perceived Ease of use (EU)}, \textit{Perceived Usefulness (U)}, \textit{Intention to Use (I)} and \textit{Correctness (C)}) using TAM model.

%Similar to other studies in the literature \cite{venigalla2020g4d,survivecovid,tlili2016improving}, we have also opted for questionnaire-based user survey approach for evaluating our game.%
For the survey, a questionnaire is created consisting of twelve statements covering all the four mentioned quality factors as shown in Table \ref{tab:questions} along with three demographic questions covering prior knowledge about ML, inquisitiveness for ML, and choice of the platform for playing online games respectively as shown in Table \ref{tab:demographics}. The current prototype version of \textit{ML-Quest} is in its initial stage and has only three-game levels, each survey questions covering a particular quality factor is applicable for all the levels. To evaluate learner's satisfaction after playing the game, we use the \textit{5-point Likert scale}, which is a quantitative evaluation method and has been commonly used for evaluation of various educational games \citep{venigalla2020g4d,survivecovid}. The learners were asked to respond to each question in the range of 1 (strongly disagree) to 5 (strongly agree).

\section{Results}

\begin{figure}[!ht]
    \centering
    \includegraphics[width = 0.7\linewidth]{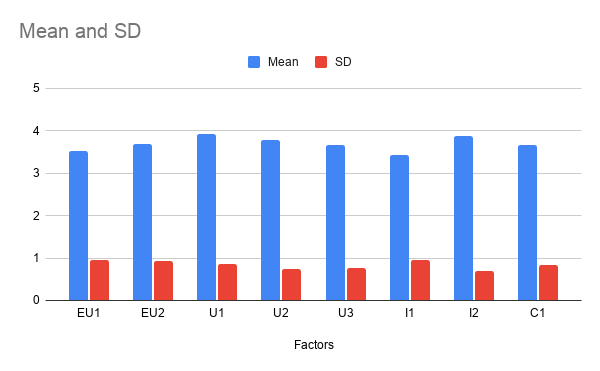}
    \caption{ Mean and Standard Deviation plots for factors of adapted TAM model in the questionnaire}
    \label{fig:msd}
    \vspace{-3mm}
\end{figure}

Twenty-three higher-secondary school students participated in the user survey, consisting of twelve questions covering the three TAM factors mentioned above, correctness and demographics. After playing the game, participants were asked to fill the questionnaire by rating each question in the range of 1 (strongly disagree) to 5 (strongly agree). A score close to 5 is an indicator of better acceptance among the users, whereas a score of 1 means lower acceptance. Figure \ref{fig:msd}, shows the result of the user survey in terms of mean and standard deviation of each question, covering all four quality factors.

According to our survey result\footnote{\url{https://osf.io/czru5/?view_only=64741df07d0e481685e2dbb2afe7a1df}} $66.7 \%$ of the participants have heard of the term ML  but only $42.1 \%$ of them searched for the underlying processes in ML. This indicates that our game could help students to build curiosity for ML and learn related concepts. The survey also reveals that most of the participants chose laptop/desktop as a platform to play games, thus encouraging us further to create such games in web-format to capture most of the audience. The demographics of the survey are presented in Table \ref{tab:demographics}.

\begin{table}
\caption{Correlation Analysis Results of adapted TAM model variables}
{\begin{tabular}{llll} \toprule
    & U      & EU     & I \\ \midrule
 U  & 1      &        &  \\
 EU & 0.2994 & 1      &  \\
 I  & 0.4952 & 0.4694 & 1 \\ \bottomrule
\end{tabular}}
\label{tab:outcomes}
\end{table}

Table \ref{tab:outcomes} shows the correlation between various factors that we considered for evaluation from the TAM model. The TAM factors I and U have a strong positive correlation indicating the fact that usage and intention are strongly dependent on each other. The main aim of the game was to introduce the concepts to children in an educative and entertaining format. As visible from Figure \ref{tab:outcomes}, question U3, ``Using ML-Quest would make it easier for me to learn Machine Learning concepts", has a mean of $3.65$ and a standard deviation of $0.775$. A mean value close to 4 with a minimum standard deviation shows that most of the participants agreed to question U3, indicating the usefulness of the game. Hence, the game can be played by interested participants to get a basic understanding of some of the concepts in ML. The levels in the game have been built with variation in complexity and an interesting storyline which makes it interactive and engaging. The results show that the participants find the storyline engaging, which can be observed from the responses of question I2, ``ML-Quest has made my learning interactive", with a high mean of $3.87$. %All these metrics are in favour of the scaffolding techniques that have been applied to teach the concepts through the game.

Correctness has a mean of $3.65$ and a standard deviation of $0.83$, which indicates the validity of our game, and users are able to relate the level outcomes to the underlying concepts of the ML concept. Overall result of the survey indicates that the game was able to capture the proposed intention, and the test audience were highly satisfied by playing the game. The game has also helped in building curiosity for the subject and tried to uncover the black-box created in today's world.
Furthermore, a few suggestions and constructive comments have also been provided by the participants such as:
\begin{itemize}
% \item  \textit{"Remembering instructions are difficult in level 1"}
\item \textit{``Player movement is too fast, needs to be slowed."}
\item \textit{``Level 2, enemies' shooting balls are not visible"}
\end{itemize}

\section{Discussion and Limitations}
%\textit{ML-Quest} is based on an RPG (Role Player Game) quest-based theme, where the story line deals with a character who needs to save herself and her town from the enemies. %The game is designed in a way that enforces the players to apply high level idea of ML concepts in simulated scenario to achieve the goal. There have been plenty of discussions on imparting \textit{Computer Science} knowledge to a younger audience at K-12 level. To augment students with supplementary knowledge it has been found useful to apply suitable abstraction for easy understanding yet necessary knowledge gain.
The developed prototype version of \textit{ML-Quest} aims to provide conceptual overview of the three ML concepts (i.e \textit{Supervised Learning}, \textit{Gradient Descent}, \textit{KNN Classification}) by motivating the player to apply those concepts in a simulated scenario. The game creates a scenario where player needs to save himself and his town members by performing the tasks at each level with the help of player's intuition and instructions provided. To make game more engaging and motivating for the player, tasks have been designed as challenges and also points are given to the player throughout the game. At each level of the game tasks have been designed aligned to the concept taught in that level and also mapped to the respective ML concept at the end of the level for better understanding of the player. Though the game has been developed to address the younger generation specifically the K-12, it can be played by a wide range of enthusiastic audience as well. %The only requirement is that the player should be equipped with a laptop and a web browser, along with a stable internet connection.

% Integrating ML to various products and services has led to growth in demand of people with skill set in ML \footnote{https://www.forbes.com/sites/insights-intelai/2019/05/22/ai-goes-to-high-school/#68826e3f1d0c}. There have been numerous efforts to improvise the college curricula along with supplementing materials to enhance the student's perception and retention of the subject. Some of these methods involve playing serious video games which has been found to effectively promote the content delivery of the subject of interest. %But as per our findings, there have been no such attempts at K-12 level in the form of games which might be of interest to the students. Our proposed game \textit{ML-Quest} aims at imparting a high level of abstraction of a few ML concepts which might help spark an interest into the field and help them explore it further and pursue it as a career.
The main challenge while designing \textit{ML-Quest} was mapping ML concepts to the gameplay and levels. A concept in ML such as supervised learning has many subsections such as the mathematical aspect, the weights, labeled test-cases, target variables, and many more. \textit{ML-Quest} has been designed focusing on the concepts which are more fundamental and uncomplicated to be put into the game. According to reported games in literature, it has been observed that a game should be interesting and engaging to its playing audience to increase the learning efficiency. For doing so, the proposed game \textit{ML-Quest} uses scaffolding technique for game designing and gradually uncover the black box as the player reaches the end of the level. Another challenge during the game development is the level of representation that needs to be provided in order for the player to understand and build a mental model of the concept. Many educational videos use extensive animations and slides, but the core aspect may still remain untouched. The game has been kept precise in terms of providing the relevant content as much as possible and evaluated using TAM model.% Still, a lot more can be further improvised.
%The game's main aim is to help players get an overview of some of the ML concepts. For this purpose, the game has various components which are abstractions of the ML concepts or some tools for understanding those concepts. 

%We have used TAM as an evaluation model for our game due to its understandability and simplicity.  
%However, it is not absolutely correct and all relationships are not necessarily and effectively observed in the studies. 
%The game has been built implementing the scaffolding technique, but the game covered only it's few aspects. The most important feature of this aspect is that scaffolding should be provided as per the student's level of comprehension. Thus, the presence of an external mentor/teacher can fulfill this purpose and increase the effectiveness by regulating it as per the student's understandings. Teachers can provide analogous examples and more useful scenarios to instantly clear relevant doubts of the players.

In the current version, \textit{ML-Quest} has only three levels to introduce the ML concepts such as \textit{Supervised Learning}, \textit{ Gradient Descent} and \textit{KNN Classification}. The game is being actively developed based on continuous feedback from players. The learning outcome is provided in a textual format which can be replaced by visual animations. The number of instructions provided in the beginning is less and could be increased for improving the experience for first-time learners. The character was limited by the asset available and can be worked upon to create custom characters for enhancing the game play.

\section{Conclusion and Future Work}
The \textit{ML-Quest} attempts at introducing the definition and working (i.e conceptual overview) of three fundamental Machine Learning (ML) concepts to school students without going into the intricate complexities.
%Considering, the popularity of games and the amount of time children spend on playing games along with the rewarding outcomes, these has been incorporated into learning various concepts such as debugging, computer architecture and so on. Still based on reported works we could not find any such game which talks about ML concepts and might target school students. We expect that our proposed game could help students to get basic understanding of high level implementation of the ML concepts in interesting way and enhance their understanding of real-world technologies. 
There are three levels in the game for teaching three different ML concepts namely \textit{Supervised learning}, \textit{Gradient Descent} and \textit{KNN Classification}. The game uses scaffolding technique to help players understand the concepts and create a mental model for the same. The storyline of the game consists of a character who needs to perform quests in order to free her kingdom. The game is engaging and consists of various components to make it interactive. \textit{ML-Quest} has been evaluated based on TAM quality factors through questionnaire based survey on 23 participants from higher-secondary school. After a remote qualitative user survey, we concluded that the majority of the learners are satisfied with the game design and agree that the game has enhanced their understanding related to ML concepts.
% The obtained survey result shows that the majority of the learners are satisfied with the game and believe that the game has cultivated their interest for the subject.
% and also 80\% of them are willing to recommend \textit{ML-Quest} to their peers.

Future works of the \textit{ML-Quest} will focus on adding more ML concepts by integrating new levels with better visualizations. We also plan to increase the complexity of the game to make it more challenging and fun at the same time by including more user engagement along with multiplayer feature. The sensitivity of the game control at each level could also be improved as suggested by our volunteers in the survey. Our survey also shows that a considerable number of students prefer mobile based games, thus we are planning to explore other platforms such as smartphone where the game can be played seamlessly.

\section*{Acknowledgement(s)}

We would like to thank all the volunteers for their valuable time and honest feedback. Special thanks to Akhila Sri Manasa Venigalla for her guidance and support from the beginning of the project and to Eashaan Rao for reviewing and making suggestions.

\bibliographystyle{ACM-Reference-Format}
\bibliography{sample-base}

\end{document}